\documentclass[preprints,article,accept,moreauthors]{Definitions/mdpi}

\firstpage{1} 
\pubvolume{xx}
\issuenum{1}
\articlenumber{5}
\pubyear{2020}
\copyrightyear{2020}
\history{Received: date; Accepted: date; Published: date}

\usepackage{hyperref}

\Title{A Simple Method to Determine Critical Coagulation Concentration from Electrophoretic Mobility}

\Author{Marco Galli $^{1,\dagger}$\orcidA{}, Szilárd Sáringer$^{2,\dagger}$\orcidB{}, István Szilágyi$^{2,}$*\orcidC{} and Gregor Trefalt $^{1,}$*\orcidD{}}

\AuthorNames{Marco Galli, Szilárd Sáringer, István Szilágyi and Gregor Trefalt}

\address{%
$^{1}$ \quad Department of Inorganic and Analytical Chemistry, University of Geneva, Sciences II, 30 Quai Ernest-Ansermet, 1205 Geneva, Switzerland; marco.galli@unige.ch (M.G.), gregor.trefalt@unige.ch (G.T.) \\
$^{2}$ \quad MTA-SZTE Lendület Biocolloids Research Group, Interdisciplinary Excellence Centre, Department of PhysicalChemistry and Materials Science, University of Szeged, 1 Rerrich Béla tér, H-6720 Szeged, Hungary; saringer.szilard@chem.u-szeged.hu (S.S.), szistvan@chem.u-szeged.hu (I.S.)}

\corres{Correspondence: gregor.trefalt@unige.ch (G.T.), szistvan@chem.u-szeged.hu (I.S.)}

\firstnote{These authors contributed equally to this work.}

\abstract{
Critical coagulation concentration (CCC) is a key parameter of particle dispersions, since it provides the threshold limit of electrolyte concentrations, above which the dispersions are destabilized due to rapid particle aggregation. A computational method is proposed to predict CCC values using solely electrophoretic mobility data without the need to measure aggregation rates of the particles. The model relies on the DLVO theory; contributions from repulsive double-layer forces and attractive van der Waals forces are included. Comparison between the calculated and previously reported experimental CCC data for the same particles shows that the method performs well in the presence of mono and multivalent electrolytes provided DLVO interparticle forces are dominant. The method is validated for particles of various compositions, shapes, and sizes.
}


\newcommand{\eq}{Eq.}
\newcommand{\eqs}{Eqs.}
\newcommand{\kkk}{K$_3$Fe(CN)$_6$}
\newcommand{\kkkk}{K$_4$Fe(CN)$_6$}

\begin{document}

\section{Introduction}
Dispersions of nano or colloidal particles attract widespread contemporary interest due their extensive use in various processes in fundamental research and in more applied disciplines, in which the phyisco-chemical properties of the particles and the media vary in a wide range~\cite{Bahng2015,Scholten2012,Xia2000a}. Applications include catalysis~\cite{Herves2012}, energy storage~\cite{Tiwari2012}, sensing~\cite{Qin2018}, drug delivery~\cite{Sokolova2008} and other biomedical utilizations~\cite{Masud2019, Gu2015}, however, the stability of these dispersions is always a key issue.
A typical example is the use of particles as catalysts in liquid media, where stable dispersions ({\sl i.e.}, homogeneously distributed primary particles) are requested during the catalytic run, while the samples can be destabilized ({\sl i.e.}, particle aggregation occurs) once the reaction is terminated~\cite{Scholten2012, Ott2007, Biondi2012}. The bigger aggregates then sediment or cream according to their density and they can be removed from the system by filtration.
Moreover, aggregation of particles must be suppressed during biomedical delivery processes, since the formation of aggregates in biofluids may cause health complications such as thrombosis by blocking the veins~\cite{Abdalla2014, Moore2015, Vasti2016}. Stable particle dispersions are also required in product manufacturing processes in the textile~\cite{Kolman2017}, food~\cite{Dickinson2010} and cosmetic~\cite{Morsella2016} industry as well as in material science, where these dispersions provide processable source of particles for building up composite materials~\cite{Guimaraes2014, Kun2005,Rouster2019,Szabo2007,Ueno2011a}.
In contrast, other applications rely on destabilization of particle dispersions by induced aggregation. A typical example is secondary water treatment, in which multivalent ions and polyelectrolytes act as aggregating agents of dispersed dust particles, which sediment and hence, can be eliminated from the tanks in the aggregated form~\cite{Bolto2007a}. Nanoparticles can also be used to eliminate toxic contaminants from waters~\cite{Simeonidis2016} and the removal of these particles also occurs by induced aggregation. Besides, the paper industry uses particles during production to improve certain properties of the paper and the particles undergo coaggregation with cellulose fibers in the papermaking process~\cite{Porubska2002a}.

It is evident from the above examples that colloidal stability of particles is a key factor in most of the applications. The usual quantity to estimate colloidal stability of a particle dispersion is the critical coagulation concentration (CCC) or critical coagulation ionic strength (CCIS) indicating the necessary electrolyte concentration for the destabilization of the sample~\cite{Elimelech1995a}. In other words, the particle collision efficiency becomes unity at the CCC, while this efficiency decreases with decreasing the concentration of the aggregating agent and becomes unmeasurable for highly stable samples, in which the particles do not form dimers after collisions.
The accurate knowledge of the CCC is therefore a critical issue to estimate colloidal stability and to design stable or unstable particle dispersions. It can be determined by measuring aggregation rates of particles with a suitable technique~\cite{Xu2011a}. One of the most handy methods involves light scattering either in static or dynamic mode~\cite{Trefalt2013c}. In brief, time-resolved measurements are applied and the change in the scattered intensity or in the hydrodynamic radius can be used to calculate aggregation rate constants. These constants were also determined in turbidity measurements by following the changes of the transmittance~\cite{Kobayashi2016} or absorbance~\cite{Gudarzi2016a} data in time-resolved experiments. A more advanced, but time-consuming approach is to measure inter-particle forces with the multi-particle colloidal probe technique based on atomic force microscopy~\cite{Sinha2013a}. From the primary experimental force-distance data, the rates can be calculated at different aggregating agent concentrations using appropriate theories and thus, the CCC can be determined.

The CCCs are typically strongly decreased in the presence of multivalent counterions. The correlation between valence of the counterion and CCC was observed more than 100 years ago by Schulze and Hardy~\cite{Hardy1899,Schulze1882}. This strong correlation is refereed to as the Schulze-Hardy rule. Later it was realized that also strongly adsorbing monovalent ions can substantially shift the CCC to lower values~\cite{Oncsik2015, Oncsik2016, Rouster2017}. Therefore, the strong interaction of the counterion with the surface is the key parameter for determining the CCC. The CCC values often correlate with the magnitude of charge or electrokinetic potential of the particles, {\sl i.e.}, particle aggregation occurs at low magnitudes of electrokinetic potentials~\cite{Franks2002a,Fernandez-Nieves1999,Trefalt2017a}. However, a threshold electrokinetic potential, under which the dispersion can be considered as unstable, cannot be generated for the individual systems, since aggregation processes depend on several factors including size of particles and ionic composition of the surrounding solution.

Here, we propose and implement a simple method for calculation of CCC, based on electrophoretic mobility. Electrophoretic mobility data (which are correlated with electrokinetic potentials~\cite{Delgado2007a}) of colloidal or nanoparticles measured at different electrolyte concentrations were used to calculate the CCC of various dispersions containing organic or inorganic charged particles of different shapes. The developed method relies on the Derjaguin, Landau, Verwey, and Overbeek (DLVO) theory, which takes into account electrostatic and van der Waals interactions. The calculated CCC values were compared to experimental ones determined independently by suitable techniques.

\section{Calculating Critical Coagulation Concentration}

In order to calculate the critical coagulation concentration (CCC), appropriate model has to be chosen. Here we employ DLVO theory for these calculations, since we are interested in aggregation of charged colloidal particles. In some systems non-DLVO interactions importantly shift the CCC, and in these cases the proposed approach is not applicable, as it will be seen in the results section. The DLVO theory assumes the interaction energy, $V$, between two charged particles to be composed of two contributions~\cite{Derjaguin1941,Verwey1948}
\begin{equation}
V = V_{\rm vdw} + V_{\rm dl} ,
\label{eq:dlvo}
\end{equation}
where $V_{\rm vdw}$ and $V_{\rm dl}$ represent the van der Waals and double-layer contributions, respectively. Van der Waals interaction between two spherical particles with radius, $R$, can be calculated as~\cite{Israelachvili2011,Russel1989} 
\begin{equation}
V_{\rm vdW} = -\frac{HR}{12}\cdot\frac{1}{h},
\label{eq:vdw}
\end{equation}
where $H$ is the Hamaker constant and $h$ is the surface separation distance. For the double-layer interaction Debye-H\"uckel superposition approximation is used
\begin{equation}
V_{\rm dl} = 2\pi R\varepsilon\varepsilon_0\psi_{\rm dl}^2 e^{-\kappa h} ,
\label{eq:dl}
\end{equation}
where $\varepsilon$ is the dielectric constant, $\varepsilon_0$ is the vacuum permittivity, $\psi_{\rm dl}$ is the diffuse-layer potential, and $\kappa$ is the inverse Debye length. The latter parameter can be calculated as
\begin{equation}
\kappa^2 = 8\pi \ell_{\rm B} N_{\rm A} 10^3 I ,
\label{eq:kappa}
\end{equation}
where $\ell_{\rm B} = \frac{e_0^2}{4\pi \varepsilon\varepsilon_0 k_{\rm B}T}$ is the Bjerrum length, $N_{\rm A}$ is the Avogadro constant, $I$ is the ionic strength expressed in (mol/L), $e_0$ is the elementary charge, $k_{\rm B}$ is the Boltzmann constant, and $T$ is the temperature. Within the Debye-H\"uckel approximation the charge density and potential of the surface are connected with
\begin{equation}
\sigma = \varepsilon\varepsilon_0 \kappa \psi_{\rm dl} .
\label{eq:charge}
\end{equation}
At CCC the repulsive energy barrier vanishes, which can be mathematically written as
\begin{equation}
V = 0 \quad {\rm and} \quad \frac{{\rm d} V}{{\rm d} h} = 0 .
\label{eq:barrier}
\end{equation}
Combining \eqs~(\ref{eq:dlvo})--(\ref{eq:charge}) with conditions in \eq~(\ref{eq:barrier}) permits to calculate the critical coagulation ionic strength (CCIS)~\cite{Trefalt2017a}
\begin{equation}
{\rm CCIS} = \frac{1}{8\pi \ell_{\rm B}} \cdot \left( \frac{24\pi}{He\varepsilon\varepsilon_0} \right)^{2/3} \sigma^{4/3} ,
\label{eq:ccis}
\end{equation}
where $e = 2.7182\dots$ is the base of the natural logarithm. \eq~(\ref{eq:ccis}) can be use to calculate the ionic strength, which corresponds to the CCC. In order to convert the CCIS to the CCC the relation between ionic strength and concentration has to be used
\begin{equation}
I = \frac{1}{2} \sum_i c_i z_i^2 ,
\label{eq:ionic_strength}
\end{equation} 
where $c_i$ is the concentration of all ionic species in the solution and $z_i$ is their valence. The relation between CCIS and the surface charge density given in \eq~(\ref{eq:ccis}) has been tested on different types of particles and different electrolytes and gives relatively accurate results~\cite{Trefalt2017a,Oncsik2015,Pavlovic2016,Rouster2017}. This analysis requires to first measure the CCC and then measure the surface charge density at the CCC. Here we propose a simple numerical procedure, which is able to predict the CCC from the measurements of electrophoretic mobility as a function of salt concentration. Electrophoretic mobility, $\mu$, is converted to electrokinetic potential, $\zeta$, via Smoluchowski equation~\cite{Russel1989}
\begin{equation}
\zeta = \frac{\mu \eta}{\varepsilon\varepsilon_0} \approx \psi_{\rm dl} ,
\label{eq:mobility}
\end{equation}
where $\eta$ is the solvent viscosity. The electrokinetic potential is usually a good approximation for the diffuse layer potential especially if the potentials are relatively low~\cite{Trefalt2016, Hartley1997}. Note that van der Waals and double-layer interaction energies written above correspond to the case of spherical colloids. As long as the dimensions of the particles are larger than the diffuse-layer thickness at the CCC, the Derjaguin approximation is valid and one can use the corresponding effective radii in the equations~\cite{Russel1989}. For small particles of arbitrary shape however some inaccuracies steaming from the use of Derjaguin approximation can be expected.

Based on the equations written above one can construct a simple algorithm to calculate the CCC from electrophoretic mobility {\sl vs} concentration data. The pseudo-code for this algorithm is given below:\\
\\
{\ttfamily
\noindent
\begin{tabular}{ll}
convert {\bf mobility} to {\bf potential} & \# use \eq~(\ref{eq:mobility})\\
convert {\bf concentration} to {\bf ionic strength} & \# use \eq~(\ref{eq:ionic_strength})\\
convert {\bf potential} to {\bf surface charge} & \# use \eq~(\ref{eq:charge})\\
calculate {\bf CCIS} from {\bf surface charge} & \# use \eq~(\ref{eq:ccis})\\
find roots for $\rm \bf (CCIS - ionic\, strength)= 0$ & \# find where calculated ccis is equal \\
 & \# to experimental ionic strength \\
\\
convert the resulting roots from {\bf CCIS} to {\bf CCC} & \# use \eq~(\ref{eq:ionic_strength})\\
\end{tabular}
}\\
\\
We have implemented this algorithm in python and the source code is available at \url{https://github.com/colloidlab/ccc-calculator}~\cite{GregorTrefalt2020}.

\section{Results and Discussion}

The data needed to calculate the critical coagulation concentration (CCC) is an array of electrophoretic mobility values with corresponding array of concentrations, electrolyte composition, viscosity of the solvent, and Hamaker constant for the measured system. All the analyzed data we show in this paper is taken from published work and the respective data sources are cited along the presented data sets.

\begin{figure}[t]
\centering
\includegraphics[width=11cm]{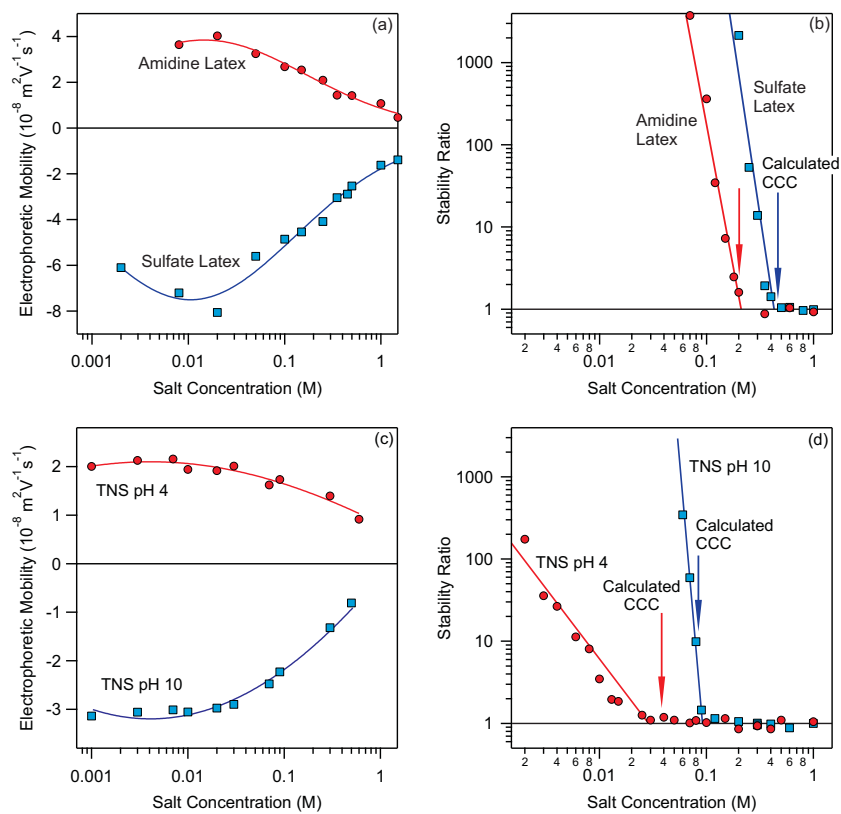}
\caption{(a) Electrophoretic mobility as a function of NaCl concentration for amidine and sulfate latex particles. The lines show the interpolating curves used for CCC calculation. (b) Stability ratios as a function of NaCl concentration for amidine and sulfate latex. Measurements were done at pH 4.0. Data were taken from~\cite{Oncsik2016}. The arrows mark the calculated CCC values based on Hamaker constant of $9.0\cdot 10^{-21}$~J. The lines connecting the stability data points are eye-guides. (c) Electrophoretic mobility as a function of NaCl concentration for titania nanosheets (TNS) at pH 4.0 and 10. The lines show the interpolating curves used for CCC calculation. (d) Stability ratios as a function NaCl concentration. Data were taken from~\cite{Saringer2019}. The arrows mark the calculated CCC values based on Hamaker constant of $1.7\cdot 10^{-20}$~J. The lines connecting the stability data points are eye-guides.}
\label{fig:details-mono}
\end{figure}

Let us first explain the procedure for calculating the CCC in more detail. The available electrophoretic mobility data as a function of concentration of added salt is first interpolated by fitting an appropriate interpolating function. Typically a function describing a log-normal distribution works well in majority of the cases. An example of such an interpolation is shown in Fig.~\ref{fig:details-mono}a, where electrophoretic mobility data is shown for positively and negatively charged latex particles in NaCl solutions.
With increasing salt concentration, the electrophoretic mobility for both particles tends to zero as the particle charge is screened by addition of salt. At low salt levels the particle suspensions are stable and the stability ratios are well above 1, see Fig.~\ref{fig:details-mono}b. At increased concentrations one can observe the transition from stable to unstable suspensions and at this transition the stability ratio reaches unity. The concentration where this transition happens is refereed to as the CCC. One can also calculate the ionic strength at this concentration using Eq.~\ref{eq:ionic_strength}, we refer to this value as critical coagulation ionic strength (CCIS). The interpolation functions shown in Fig.~\ref{fig:details-mono} are used to calculate the CCCs by the procedure described above. The resulting calculated CCCs for amidine and sulfate latex particles are, 225~mM and 460~mM respectively. These values are marked with arrows in Fig.~\ref{fig:details-mono}b and match perfectly with the experimentally measured stability ratios.

In Fig.~\ref{fig:details-mono}cd results for titania nanosheets at two different pH are shown~\cite{Saringer2019}. These particles are postively charged at pH 4.0 and negatively charged at pH 10. Again the calculated CCCs marked with arrows in Fig.~\ref{fig:details-mono}d match perfectly the experimental data. The examples shown in Fig.~\ref{fig:details-mono} confirm that our method for calculating CCCs gives good results for positively and negatively charged as well as organic and inorganic particles in the presence of simple monovalent electrolytes.

\begin{figure}[t]
\centering
\includegraphics[width=11cm]{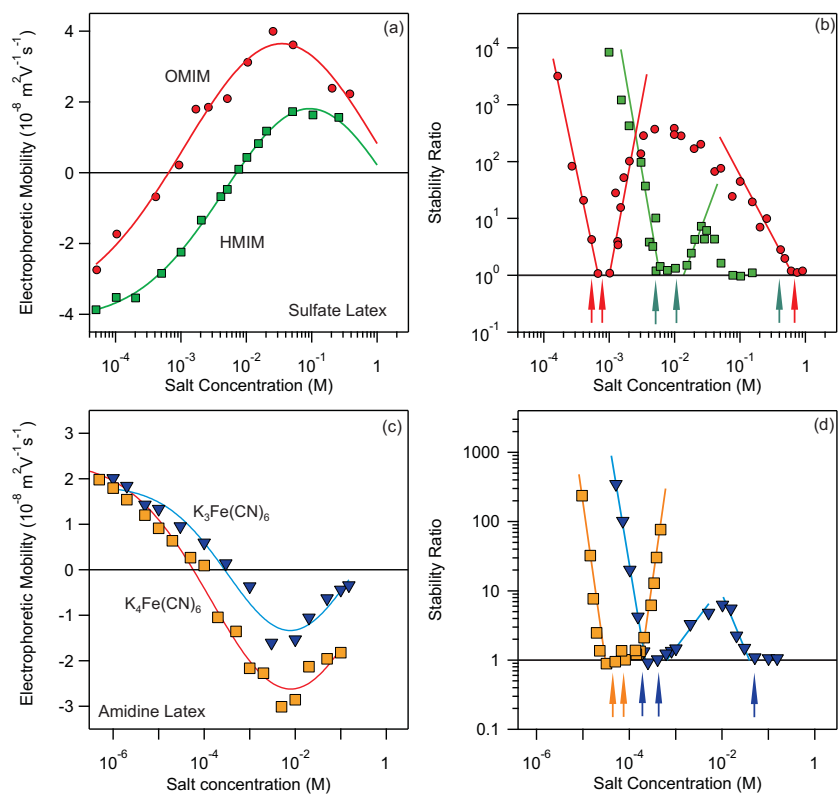}
\caption{(a) Electrophoretic mobility as a function of HMIMCl and OMIMCl concentration for sulfate latex particles. The lines show the interpolating curves used for CCC calculation. (b) Stability ratios as a function of HMIMCl and OMIMCl concentration for sulfate latex. Measurements were done at pH 4.0. Data were taken from~\cite{Oncsik2016}. The arrows mark the calculated CCC values based on Hamaker constant of $2.0\cdot 10^{-21}$~J. The lines connecting the stability data points are eye-guides. (c) Electrophoretic mobility as a function of \kkk\ and \kkkk\ concentration for amidine latex particles. The lines show the interpolating curves used for CCC calculation. (d) Stability ratios as a function of \kkk\ and \kkkk\ concentration for sulfate latex. Measurements were done at pH 4.0. Data were taken from~\cite{Cao2017}. The arrows mark the calculated CCC values based on Hamaker constant of $3.0\cdot 10^{-21}$~J. The lines connecting the stability data points are eye-guides.}
\label{fig:details-multi}
\end{figure}

Let us now look at examples, where monovalent and multivalent ions strongly adsorb to particles surfaces. In the first case, shown in the top panel of Fig.~\ref{fig:details-multi}, negatively charged sulfate latex particles are in contact with 1-hexyl-3-methylimidazolium (HMIM$^+$) or 1-octyl-3-methylimidazolium (OMIM$^+$) ions.
The particles are negatively charged at low concentrations and undergo charge-reversal at increased concentrations. This strong adsorption of imidazolium based ions is reflected also in stability of suspensions, see Fig.~\ref{fig:details-multi}b. The particles are stable at low concentration and aggregate fast close to charge-neutralization point after which they re-stabilize due to overcharging and finally become unstable again at high salt levels. This behavior gives rise to three CCCs. Our method can predict three CCCs for each salt and they are marked with red and green arrows, respectively. Similarly, the charge reversal is observed for amidine latex particles in the presence of Fe(CN)$_6^{3-}$ and Fe(CN)$_6^{4-}$ multivalent anions, see Fig.~\ref{fig:details-multi}c. Again multiple CCCs are present and can be predicted relatively accurately with the proposed method (see positions of the arrows in Fig.~\ref{fig:details-multi}d). However, the accuracy of predictions for strongly adsorbing ions is in some cases lower. For example, in the case of HMIMCl the third CCC is predicted at $\sim 400$~mM while it is observed below 100~mM. These deviations are probably related to non-DLVO interactions, which are not taken into account in our simple method, however they have been shown to be important in similar systems~\cite{Moazzami-Gudarzi2018,Smith2018,Cao2017}. Albeit some inaccuracies, our proposed method for calculating CCCs is still able to quantitatively predict CCCs in most cases.

\begin{figure}[t]
\centering
\includegraphics[width=15cm]{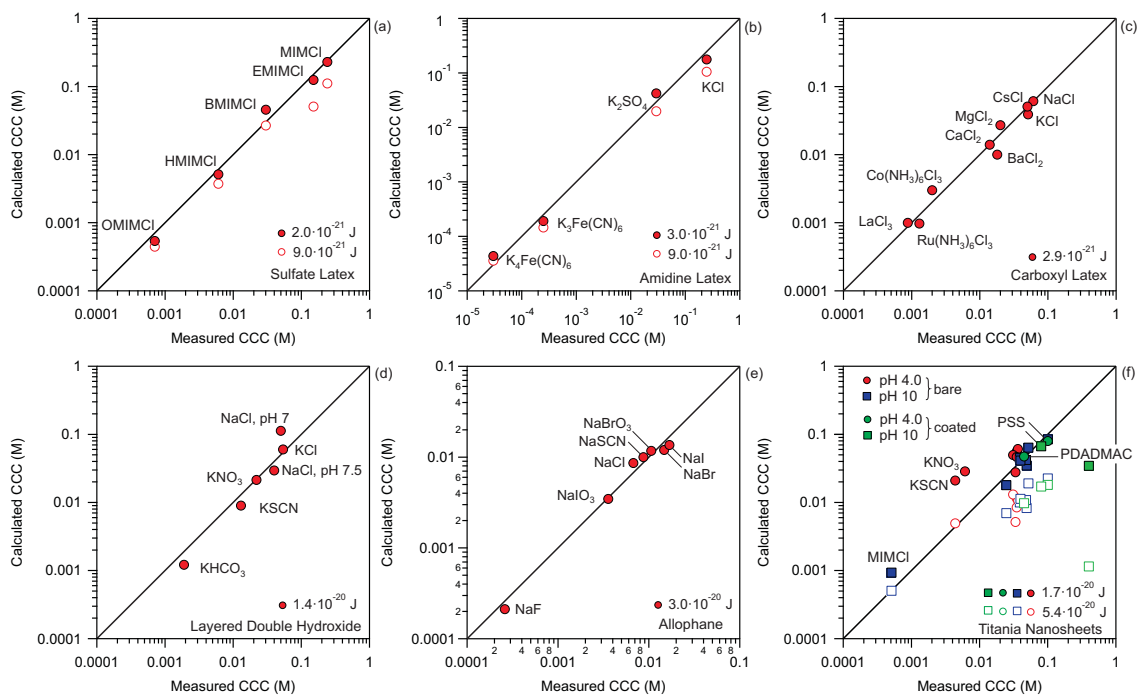}
\caption{Calculated versus measured CCCs for different systems. Full symbols adjusted Hamaker constant, empty symbols theoretical Hamaker constants. (a) Sulfate latex particles in the presence of imidazolium based salts~\cite{Oncsik2016}. (b) Amidine latex particles in the presence of multivalent anions~\cite{Cao2017}. (c) Carboxyl latex particles in the presence of mono and multivalent cations~\cite{Oncsik2014a}. (d) Layered double hydroxide platelets in the presence of monovalent salts~\cite{Pavlovic2016}. (e) Allophane clay nanoparticles in the presence of monovalent anions~\cite{Takeshita2019}. (f) Bare and coated titania nanosheets in the presence of monovalent satls~\cite{Rouster2019a, Saringer2019, Rouster2017}. Theoretical values of Hamaker constants for polystyrene latex particles and titania particles of $9.0\cdot 10^{-21}$~\cite{Elzbieciak-Wodka2014} and $5.4\cdot 10^{-20}$~\cite{Bergstrom1997} are used, respectively. All the points in the figure are listed in Table~\ref{tab:systems}.}
\label{fig:calc_measured}
\end{figure}

In the following we focus more on quantitative aspects of the proposed method. In Fig.~\ref{fig:calc_measured} calculated versus measured CCCs are shown for seven different types of particles at different conditions.
This analysis enables us to judge the quantitative accuracy of the proposed method. The results for different types of organic and inorganic particles as well as spherical and layered materials are shown. For each system CCCs resulting from the aggregation induced by different ions are shown. All the calculations for a specific system are done with one value for Hamaker constant. These Hamaker constants need to be adjusted because theoretical values of Hamaker constants are either not known or they can overestimate the magnitude of the van der Waals force. The Hamaker constants used for polystyrene latex particles shown in Fig.~\ref{fig:calc_measured}a-c are between 2 and $3\cdot 10^{-21}$~J. These values are in agreement with typical values for latex particles in aqueous solutions extracted from direct force measurements~\cite{Moazzami-Gudarzi2018, Smith2018, Elzbieciak-Wodka2014}. Note that however, the measured values are substantially lower as compared to the theoretical value for polystyrene latex calculated from the Liftsitz theory, which is equal to $9.0\cdot 10^{-21}$~J~\cite{Elzbieciak-Wodka2014}. These deviations between the measured and calculated Hamaker constants can be explained by the surface roughness of the particles used in the experiments~\cite{Elzbieciak-Wodka2014, Valmacco2016, Thormann2017}. Therefore, the use of theoretical values of the Hamaker constants for calculation of CCCs leads to underestimation of CCCs, see Fig.~\ref{fig:calc_measured}a, b, and f. This procedure of adjusting the Hamaker constants for a given material therefore enables the determination of the effective Hamaker constant for the system at hand.

In the majority of the cases the points shown in Fig.~\ref{fig:calc_measured} lay very close to the diagonal line. These results confirm that our simple method is able to quantitatively predict the CCCs for a wide variety of systems. These systems include spherical, non-spherical, and platelet particles composed of different materials in the presence of simple monovalent as well as multivalent, and complex organic ions. In Fig.~\ref{fig:calc_measured}f also results for coated titania nanosheets are shown. In the presented case the titania nanosheets were first coated with poly(diallyldimethylammonium chloride) (PDADMAC) or poly(styrene sulfonate) (PSS) polyelectrolytes and then their stability was measured as a function of NaCl concentration at pH 4.0 and pH 10~\cite{Saringer2019}. Again one can quantitatively predict the CCCs of polyelectrolyte coated particles with the Hamaker constant of $1.7\cdot 10^{-20}$~J, which is also used for uncoated titania nanosheets. A more detailed observation of the results shown in Fig.~\ref{fig:calc_measured} reveals that in certain cases marked deviations between calculated and measured CCCs exist. The most prominent deviation is visible in Fig.~\ref{fig:calc_measured}f for titania nanosheets coated with PDADMAC at pH 10. For this system the calculation underestimates the CCC by almost an order of magnitude. In other words, the experimental system is more stable than predicted. This result is probably a consequence of some additional steric repulsion between two platelets induced by adsorbed polyelectrolytes, which is not part of our DLVO model. Other deviations include titania particles in the presence of KSCN and KNO$_3$ salts at pH 4.0. In these cases non-DLVO interactions are probably present due to strong specific interactions of anions with the surface. In general, for the systems below the diagonal line, non-DLVO repulsions, while for the systems above this line, non-DLVO attractions are present. Therefore, these deviations can give us a further information about the presence of the non-DLVO forces.

\begin{figure}[t]
\centering
\includegraphics[width=15cm]{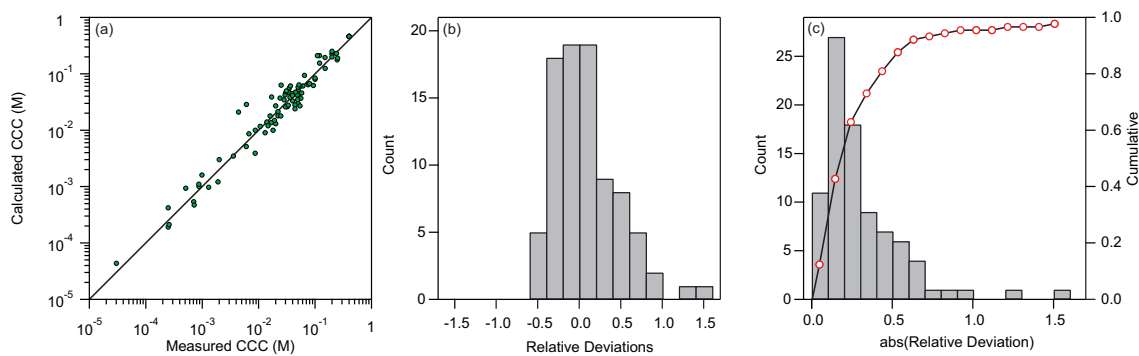}
\caption{Deviations of calculated and experimental CCCs. (a) Comparison of calculated and measured CCCs for all samples. (b) Distribution of relative deviations between calculated and measured CCCs. (c) Distribution of absolute values of relative deviations between calculated and measured CCCs shown with bars with scale on the left side. The points show the cumulative of the distribution with scale on the right side.}
\label{fig:deviations}
\end{figure}

Finally we perform some statistical analysis on the performance of the proposed calculation method. In addition to the systems shown in Fig.~\ref{fig:calc_measured} additional data points were collected and are shown in Fig.~\ref{fig:deviations}a and in Table~\ref{tab:systems}.
The data in Fig.~\ref{fig:deviations}a show that the CCCs can be predicted relatively accurately over more than four orders of magnitude. The distribution of the relative deviation of the calculated CCCs from the measured CCCs is shown Fig.~\ref{fig:deviations}b. Vast majority of the samples fall between the deviations of [$-40~\%$, 40~\%]. Few samples have deviations above 100~\% and for these samples additional non-DLVO interactions substantially shift the CCC. Furthermore, the deviations in the positive side are more prominent. This observation can be explained by the fact that in majority of cases non-DLVO interactions, which considerably contribute to the shifting of the CCC, are attractive. Such attractive non-DLVO interactions are typically observed in systems, where ions strongly interact with the surface~\cite{Moazzami-Gudarzi2018,Smith2018,Cao2017}.

By taking the absolute values of the relative deviations one can construct a distribution of deviations shown in Fig.~\ref{fig:deviations}c. The average deviation for all the sample is 38~\%. However, this relatively big average deviation is due to the skewed distribution. There are few samples, for which the deviation is extremely big and these points have a big influence on the average deviation. For more than 92~\% of the points the deviation is smaller that 0.75. For the points with the deviations larger than 0.75 the influence of the non-DLVO interactions is substantial and our proposed method is not applicable. If these points are omitted, the average deviation comes down to 25~\%. Bearing in mind that the error of the measurements of the CCCs is typically between 10-20~\% one can conclude that our proposed method performs surprisingly well.

\begin{figure}[t]
\centering
\includegraphics[width=11cm]{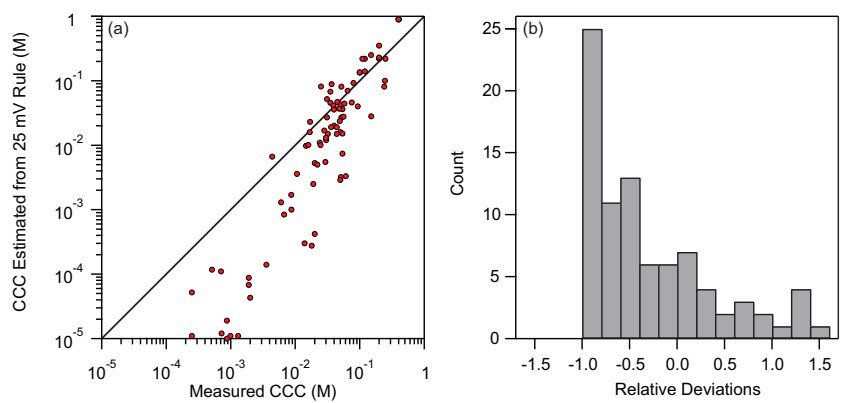}
\caption{(a) Deviations of CCC estimated using the 25~mV rule and experimental CCCs. (b) Distribution of relative deviations between estimated CCCs using the 25~mV rule and measured CCCs.}
\label{fig:deviations-25}
\end{figure}

Let us finally address the conventional wisdom that states that the suspensions lose their stability when the electrokinetic potential of particles reduces below 25~mV~\cite{Vallar1999}. As it was already shown by some of us~\cite{Trefalt2017a} this rule is not applicable for suspensions containing multivalent ions. Here we are further testing this simple rule by estimating the CCCs using 25~mV as a threshold of stability. The deviations between CCCs estimated from 25~mV rule and experimental CCCs are shown in Fig.~\ref{fig:deviations-25}.
While the 25~mV rule works reasonably well for the CCCs between 0.1 and 1~M, the results are significantly deviating for lower concentrations. This systematic deviation is clearly evident from the fact that the points in Fig.~\ref{fig:deviations-25}a follow a slope steeper than one. Furthermore, the histogram of the relative deviations presented in Fig.~\ref{fig:deviations-25}b shows that the majority of points have pronounced negative deviations. The average relative deviations in this case is larger than 65~\%, which is much worse than our proposed method. What is even more dramatic is the systematic shift to negative deviations. The 25~mV rule is clearly unreliable for low CCCs and therefore its applicability is very limited.

\section{Conclusions}

In the present work, experimental electrophoretic mobility values measured in electrolyte solutions were used to calculate CCC of dispersed particles without direct determination of their aggregation rates. The developed method is based on the DLVO theory, {\sl i.e.}, charged particles are expected to form stable dispersions at low electrolyte concentrations due to the stabilization effect by the electrical double-layer force and the samples are destabilized above the CCC by the van der Waals attractions. For systems where the experimental Hamaker constant is not known, this parameter needs to be adjusted in order to get reliable estimations of the CCC. Therefore this method can also be used for determination of the effective Hamaker constant by measuring CCCs with different salts for the same particle system and comparing them to the calculated values.

Statistical analysis of the results and comparison to experimental CCC data determined by independent measurements revealed, that the implemented method provides CCCs, which agree very well with the experimental ones. Note that presence of non-DLVO forces may lead to certain failure of the method, as pointed out with polymer-coated particles, where steric forces provided additional stabilization of the dispersions giving rise to deviation between the calculated and experimental CCC. The method was tested for various systems including particles of different composition (polymer, metal hydroxide, and oxide), structure (sphere, sheet, and lamellar) and size (nano and colloidal particles). In addition, several types of salt solutions (ionic liquid constituents, mono and multivalent electrolytes) were used as dispersion medium. The developed method is able to determine CCC values with a deviation from the experimental values of less than 25~\% for majority of samples. On the other hand a simple rule based on the assumption that the CCC occurs when electrokinetic potential reaches 25~mV is much less accurate and has large systematic deviations for small CCC values, which suggests that this assumption is not physically sound. 

 The developed algorithm is provided as free and open source software, allowing other researchers to determine CCCs solely from electrophoretic mobility data. This tool can be of special importance in systems, in which the direct measurement of the CCC is not possible due to non-ideal sample conditions (e.g., high polydispersity, size and concentration of the particles) and also in dispersions containing electrolyte mixtures, as in many industrial and environmental processes.

\vspace{6pt} 



\authorcontributions{Conceptualization, I.S. and G.T.; methodology, M.G., S.S., I.S. and G.T.; software, G.T.; validation, M.G., S.S., I.S. and G.T.; resources, M.G., S.S., I.S. and G.T.; data curation, M.G., S.S., I.S. and G.T.; writing--original draft preparation, I.S. and G.T.; writing--review and editing, M.G., S.S., I.S. and G.T.; visualization, M.G., S.S., I.S. and G.T.; supervision, I.S. and G.T.; funding acquisition, I.S. and G.T. All authors have read and agreed to the published version of the manuscript.}

\funding{Financial support by the Swiss National Science Foundation through grant 162420 and National Research, Development and Innovation Office (SNN131558) and the Ministry of Human Capacities of Hungary (20391-3/2018/FEKUSTRAT) is gratefully acknowledged.}

\conflictsofinterest{The authors declare no conflict of interest.}

\appendixtitles{no}
\appendix
\section{}
\unskip

\begin{table}[H]
\caption{Table with details of the systems analyzed in the present paper. All the calculated and measured CCCs are given.}
\centering
\tablesize{\footnotesize}
\begin{tabular}{llllllc}
\toprule
\textbf{Particle}	& \textbf{Salt}	& \textbf{pH} & \textbf{Hamaker} & \textbf{Measured CCC (M)} & \textbf{Calculated CCC (M)} & \textbf{Reference}\\
	& 	&  & \textbf{Constant (J)} &  &  & \\
\midrule
Sulfate Latex      & NaCl              & 4.0 &$9.0\cdot10^{-21}$   & 0.12        & 0.21       & \cite{Oncsik2014a} \\
Sulfate Latex      & KCl               & 4.0 &$9.0\cdot10^{-21}$   & 0.11        & 0.21       & \cite{Oncsik2014a} \\
Sulfate Latex      & CsCl              & 4.0 &$9.0\cdot10^{-21}$   & 0.25        & 0.19       & \cite{Oncsik2014a} \\
Sulfate Latex      & MgCl$_2$             & 4.0 &$9.0\cdot10^{-21}$   & 0.031       & 0.048      & \cite{Oncsik2014a} \\
Sulfate Latex      & CaCl$_2$             & 4.0 &$9.0\cdot10^{-21}$   & 0.032       & 0.026      & \cite{Oncsik2014a} \\
Sulfate Latex      & BaCl$_2$             & 4.0 &$9.0\cdot10^{-21}$   & 0.024       & 0.037      & \cite{Oncsik2014a} \\
Sulfate Latex      & LaCl$_3$             & 4.0 &$9.0\cdot10^{-21}$   & 0.00099     & 0.0016     & \cite{Oncsik2014a} \\
Sulfate Latex      & Co(NH$_3$)$_6$Cl$_3$ & 4.0 &$9.0\cdot10^{-21}$   & 0.00087     & 0.0011     & \cite{Oncsik2014a} \\
Sulfate Latex      & Ru(NH$_3$)$_6$Cl$_3$ & 4.0 &$9.0\cdot10^{-21}$   & 0.00072     & 0.00047    & \cite{Oncsik2014a} \\
Carboxyl Latex & NaCl              & 4.0 & $2.9\cdot10^{-21}$ & 0.061       & 0.061     & \cite{Oncsik2014a} \\
Carboxyl Latex & KCl               & 4.0 & $2.9\cdot10^{-21}$ & 0.051       & 0.039      & \cite{Oncsik2014a} \\
Carboxyl Latex & CsCl              & 4.0 & $2.9\cdot10^{-21}$ & 0.050        & 0.051     & \cite{Oncsik2014a} \\
Carboxyl Latex & MgCl$_2$             & 4.0 & $2.9\cdot10^{-21}$ & 0.020        & 0.027      & \cite{Oncsik2014a} \\
Carboxyl Latex & CaCl$_2$             & 4.0 & $2.9\cdot10^{-21}$ & 0.014       & 0.014      & \cite{Oncsik2014a} \\
Carboxyl Latex & BaCl$_2$             & 4.0 & $2.9\cdot10^{-21}$ & 0.018       & 0.010       & \cite{Oncsik2014a} \\
Carboxyl Latex & LaCl$_3$             & 4.0 & $2.9\cdot10^{-21}$ & 0.00088     & 0.0010      & \cite{Oncsik2014a} \\
Carboxyl Latex & Co(NH$_3$)$_6$Cl$_3$ & 4.0 & $2.9\cdot10^{-21}$ & 0.0020       & 0.0030      & \cite{Oncsik2014a} \\
Carboxyl Latex & Ru(NH$_3$)$_6$Cl$_3$ & 4.0 & $2.9\cdot10^{-21}$ & 0.0013      & 0.00097    & \cite{Oncsik2014a} \\
Amidine Latex      & NaCl              & 4.0 &$9.0\cdot10^{-21}$   & 0.20         & 0.23      & \cite{Oncsik2016} \\
Amidine Latex      & NaBr              & 4.0 &$9.0\cdot10^{-21}$   & 0.12        & 0.155      & \cite{Oncsik2016} \\
Amidine Latex      & NaN(CN)$_2$          & 4.0 &$9.0\cdot10^{-21}$   & 0.050        & 0.030       & \cite{Oncsik2016} \\
Amidine Latex      & NaSCN             & 4.0 &$9.0\cdot10^{-21}$   & 0.052       & 0.044      & \cite{Oncsik2016} \\
Amidine Latex      & BMIMCl            & 4.0 &$9.0\cdot10^{-21}$   & 0.20         & 0.25       & \cite{Oncsik2016} \\
Amidine Latex      & BMIMBr            & 4.0 &$9.0\cdot10^{-21}$   & 0.15        & 0.194      & \cite{Oncsik2016} \\
Amidine Latex      & BMIMN(CN)$_2$        & 4.0 &$9.0\cdot10^{-21}$   & 0.075       & 0.064      & \cite{Oncsik2016} \\
Amidine Latex      & BMIMSCN           & 4.0 &$9.0\cdot10^{-21}$   & 0.020        & 0.013      & \cite{Oncsik2016} \\
Amidine Latex      & BMPLCl            & 4.0 &$9.0\cdot10^{-21}$   & 0.20         & 0.20        & \cite{Oncsik2016} \\
Amidine Latex      & BMPLBr            & 4.0 &$9.0\cdot10^{-21}$   & 0.065       & 0.094      & \cite{Oncsik2016} \\
Amidine Latex      & BMPLN(CN)$_2$        & 4.0 &$9.0\cdot10^{-21}$   & 0.050        & 0.058      & \cite{Oncsik2016} \\
Amidine Latex      & BMPLSCN           & 4.0 &$9.0\cdot10^{-21}$   & 0.040        & 0.033      & \cite{Oncsik2016} \\
Sulfate Latex      & NaCl              & 4.0 &$9.0\cdot10^{-21}$   & 0.40         & 0.46       & \cite{Oncsik2016} \\
Sulfate Latex      & NaBr              & 4.0 &$9.0\cdot10^{-21}$   & 0.40         & 0.46       & \cite{Oncsik2016} \\
Sulfate Latex      & NaN(CN)$_2$          & 4.0 &$9.0\cdot10^{-21}$   & 0.40         & 0.46       & \cite{Oncsik2016} \\
Sulfate Latex      & NaSCN             & 4.0 &$9.0\cdot10^{-21}$   & 0.40         & 0.46       & \cite{Oncsik2016} \\
Sulfate Latex      & BMIMCl            & 4.0 &$9.0\cdot10^{-21}$   & 0.030        & 0.026      & \cite{Oncsik2016} \\
Sulfate Latex      & BMIMBr            & 4.0 &$9.0\cdot10^{-21}$   & 0.019       & 0.015      & \cite{Oncsik2016} \\
Sulfate Latex      & BMIMN(CN)$_2$        & 4.0 &$9.0\cdot10^{-21}$   & 0.036       & 0.038      & \cite{Oncsik2016} \\
Sulfate Latex      & BMIMSCN           & 4.0 &$9.0\cdot10^{-21}$   & 0.093       & 0.062      & \cite{Oncsik2016} \\
Sulfate Latex      & BMPLCl            & 4.0 &$9.0\cdot10^{-21}$   & 0.044       & 0.028      & \cite{Oncsik2016} \\
Sulfate Latex      & BMPLBr            & 4.0 &$9.0\cdot10^{-21}$   & 0.044       & 0.024      & \cite{Oncsik2016} \\
Sulfate Latex      & BMPLN(CN)$_2$        & 4.0 &$9.0\cdot10^{-21}$   & 0.022       & 0.018      & \cite{Oncsik2016} \\
Sulfate Latex      & BMPLSCN           & 4.0 &$9.0\cdot10^{-21}$   & 0.0087      & 0.0039     & \cite{Oncsik2016} \\
Sulfate Latex      & MIMCl               & 4.0 & $2.0\cdot10^{-21}$  & 0.24     & 0.23      & \cite{Oncsik2016} \\
Sulfate Latex      & EMIMCl              & 4.0 & $2.0\cdot10^{-21}$  & 0.151    & 0.125      & \cite{Oncsik2016} \\
Sulfate Latex      & BMIMCl              & 4.0 & $2.0\cdot10^{-21}$  & 0.030    & 0.046     & \cite{Oncsik2016} \\
Sulfate Latex      & HMIMCl              & 4.0 & $2.0\cdot10^{-21}$  & 0.0061   & 0.0051    & \cite{Oncsik2016} \\
Sulfate Latex      & OMIMCl              & 4.0 & $2.0\cdot10^{-21}$  & 0.00071    & 0.00054   & \cite{Oncsik2016} \\
Amidine Latex      & KCl               & 4.0 & $3.0\cdot10^{-21}$   & 0.25     & 0.18      & \cite{Cao2017} \\
Amidine Latex      & K$_2$SO$_4$             & 4.0 & $3.0\cdot10^{-21}$   & 0.029   & 0.042     & \cite{Cao2017} \\
Amidine Latex      & K$_3$Fe(CN)$_6$         & 4.0 & $3.0\cdot10^{-21}$   & 0.00025   & 0.00019   & \cite{Cao2017} \\
Amidine Latex      & K$_4$Fe(CN)$_6$         & 4.0 & $3.0\cdot10^{-21}$   & 0.000030 & 0.000044   & \cite{Cao2017} \\
\bottomrule
\end{tabular}
\label{tab:systems}
\end{table}

\begin{table}[H]
\centering
\tablesize{\footnotesize}
\begin{tabular}{llllllc}
\toprule
\textbf{Particle}	& \textbf{Salt}	& \textbf{pH} & \textbf{Hamaker} & \textbf{Measured CCC (M)} & \textbf{Calculated CCC (M)} & \textbf{Reference}\\
	& 	&  & \textbf{Constant (J)} &  &  & \\
\midrule
Allophane          & NaF                & 5 & $3.0\cdot10^{-20}$   & 0.00026     & 0.00021   & \cite{Takeshita2019} \\
Allophane          & NaCl               & 5 & $3.0\cdot10^{-20}$   & 0.0068     & 0.0086    & \cite{Takeshita2019} \\
Allophane          & NaBr               & 5 & $3.0\cdot10^{-20}$   & 0.015      & 0.012      & \cite{Takeshita2019} \\
Allophane          & NaI                & 5 & $3.0\cdot10^{-20}$   & 0.017      & 0.0136     & \cite{Takeshita2019} \\
Allophane          & NaBrO$_3$             & 5 & $3.0\cdot10^{-20}$   & 0.0106      & 0.0117     & \cite{Takeshita2019} \\
Allophane          & NaIO$_3$              & 5 & $3.0\cdot10^{-20}$   & 0.0036     & 0.0035    & \cite{Takeshita2019} \\
Allophane          & NaSCN              & 5 & $3.0\cdot10^{-20}$   & 0.0087     & 0.010       & \cite{Takeshita2019} \\
LDH & KCl               & 9 & $1.4\cdot10^{-20}$ & 0.054       & 0.060 & \cite{Pavlovic2016} \\
LDH & KNO$_3$              & 9 & $1.4\cdot10^{-20}$ & 0.022       & 0.021  & \cite{Pavlovic2016} \\
LDH & KSCN              & 9 & $1.4\cdot10^{-20}$ & 0.013       & 0.0090 & \cite{Pavlovic2016} \\
LDH & KHCO$_3$             & 9 & $1.4\cdot10^{-20}$ & 0.0019      & 0.0012 & \cite{Pavlovic2016} \\
TNP & KCl    & 10 & $1.7\cdot10^{-20}$ & 0.025 & 0.063 & \cite{Rouster2019a}     \\
TNP & MIMCl  & 10 & $1.7\cdot10^{-20}$ & 0.00025 & 0.00042 & \cite{Rouster2019a}     \\
TNP & EMIMCl & 10 & $1.7\cdot10^{-20}$ & 0.016 & 0.018 & \cite{Rouster2019a}     \\
TNP & BMIMCI & 10 & $1.7\cdot10^{-20}$ & 0.028 & 0.027 & \cite{Rouster2019a}     \\
TNP & KCl    & 4.0  & $1.7\cdot10^{-20}$ & 0.058 & 0.046 & \cite{Rouster2019a}     \\
TNP & MIMCl  & 4.0  & $1.7\cdot10^{-20}$ & 0.056 & 0.037 & \cite{Rouster2019a}     \\
TNP & EMIMCl & 4.0  & $1.7\cdot10^{-20}$ & 0.054 & 0.027 & \cite{Rouster2019a}     \\
TNP & BMIMCI & 4.0  & $1.7\cdot10^{-20}$ & 0.040 & 0.042 & \cite{Rouster2019a}     \\
TNS    & KCl    & 10 & $1.7\cdot10^{-20}$ & 0.048 & 0.042 & \cite{Rouster2019a}     \\
TNS    & MIMCl  & 10 & $1.7\cdot10^{-20}$ & 0.00051 & 0.00094 & \cite{Rouster2019a}     \\
TNS    & EMIMCl & 10 & $1.7\cdot10^{-20}$ & 0.025 & 0.018 & \cite{Rouster2019a}     \\
TNS    & BMIMCI & 10 & $1.7\cdot10^{-20}$ & 0.049 & 0.035 & \cite{Rouster2019a}     \\
TNS    & KCl    & 4.0  & $1.7\cdot10^{-20}$ & 0.035 & 0.047 & \cite{Rouster2019a}     \\
TNS    & MIMCl  & 4.0  & $1.7\cdot10^{-20}$ & 0.035 & 0.056 & \cite{Rouster2019a}     \\
TNS    & EMIMCl & 4.0  & $1.7\cdot10^{-20}$ & 0.037 & 0.061 & \cite{Rouster2019a}     \\
TNS    & BMIMCI & 4.0  & $1.7\cdot10^{-20}$ & 0.031 & 0.050 & \cite{Rouster2019a}     \\
TNS    & NaCl   & 4.0  & $1.7\cdot10^{-20}$ & 0.017 & 0.039 & \cite{Saringer2019} \\
TNS, PDADMAC coated & NaCl   & 4.0  & $1.7\cdot10^{-20}$ & 0.045 & 0.047 & \cite{Saringer2019} \\
TNS, PSS coated & NaCl   & 4.0  & $1.7\cdot10^{-20}$ & 0.100 & 0.080 & \cite{Saringer2019} \\
TNS & NaCl   & 10  & $1.7\cdot10^{-20}$ & 0.10 & 0.084 & \cite{Saringer2019} \\
TNS, PDADMAC coated & NaCl   & 10 & $1.7\cdot10^{-20}$ & 0.40 & 0.034 & \cite{Saringer2019} \\
TNS, PSS coated & NaCl   & 10 & $1.7\cdot10^{-20}$ & 0.080 & 0.067 & \cite{Saringer2019} \\
TNS    & KCl    & 4.0  & $1.7\cdot10^{-20}$ & 0.034 & 0.028 & \cite{Rouster2017}    \\
TNS    & KNO$_3$   & 4.0  & $1.7\cdot10^{-20}$ & 0.0061 & 0.029 & \cite{Rouster2017}    \\
TNS    & KSCN   & 4.0  & $1.7\cdot10^{-20}$ & 0.0044 & 0.021 & \cite{Rouster2017}    \\
TNS    & KCl    & 10 & $1.7\cdot10^{-20}$ & 0.039 & 0.045 & \cite{Rouster2017}    \\
TNS    & KNO$_3$   & 10 & $1.7\cdot10^{-20}$ & 0.040 & 0.041 & \cite{Rouster2017}    \\
TNS    & KSCN   & 10 & $1.7\cdot10^{-20}$ & 0.052 & 0.064 & \cite{Rouster2017}   \\
\bottomrule
\end{tabular}
\end{table}
{\footnotesize\noindent Acronyms used in the table: LDH: layered double hydroxide, TNP: titania nanoparticles, TNS: titania nanosheets, PDADMAC: poly(diallyldimethylammonium chloride), PSS: poly(styrene sulfonate), BMPL: 1-butyl-1-methylpyrrolidinium, MIM: 3-methylimidazolium, EMIM: 1-ethyl-3-methylimidazolium, BMIM: 1-butyl-3-methylimidazolium, HMIM: 1-hexyl-3-methylimidazolium, OMIM: 1-octyl-3-methylimidazolium. 
}

\reftitle{References}

\externalbibliography{yes}
\bibliography{calculating_ccc.bib}

\end{document}